# Electronic structure of the Co(0001)/MoS$_2$ interface, and its possible use for electrical spin injection in a single MoS$_2$ layer

T. Garandel[1,2], R. Arras[1], X. Marie[2], P. Renucci[2], and L. Calmels[1]

[1] *CEMES-CNRS, Université de Toulouse, 29 rue Jeanne Marvig, BP 94347, F-31055 Toulouse, France*
[2] *LPCNO, Université de Toulouse, CNRS, INSA, UPS, 135 Avenue de Rangueil, 31077 Toulouse, France*

**ABSTRACT**

The ability to perform efficient electrical spin injection from ferromagnetic metals into two-dimensional semiconductor crystals based on transition metal dichalcogenide monolayers is a prerequisite for spintronic and valleytronic devices using these materials. Here, the *hcp* Co(0001)/MoS$_2$ interface electronic structure is investigated by first-principles calculations based on the density functional theory. In the lowest energy configuration of the hybrid system after optimization of the atomic coordinates, we show that interface sulfur atoms are covalently bound to one, two or three cobalt atoms. A decrease of the Co atom spin magnetic moment is observed at the interface, together with a small magnetization of S atoms. Mo atoms also hold small magnetic moments which can take positive as well as negative values. The charge transfers due to covalent bonding between S and Co atoms at the interface have been calculated for majority and minority spin electrons and the connections between these interface charge transfers and the induced magnetic properties of the MoS$_2$ layer are discussed. Band structure and density of states of the hybrid system are calculated for minority and majority spin electrons, taking into account spin-orbit coupling. We demonstrate that MoS$_2$ bound to the Co contact becomes metallic due to hybridization between Co *d* and S *p* orbitals. For this metallic phase of MoS$_2$, a spin polarization at the Fermi level of 16 % in absolute value is calculated, that could allow spin injection into the semiconducting MoS$_2$ monolayer channel. Finally, the symmetry of the majority and minority spin electron wave functions at the Fermi level in the Co-bound metallic phase of MoS$_2$ and the orientation of the border between the metallic and semiconducting phases of MoS$_2$ are investigated, and their impact on spin injection into the MoS$_2$ channel is discussed.

PACS Numbers: 72.25.Dc, 72.25.Hg, 72.25.Mk, 73.20.At, 73.63.Rt, 75.70.Ak, 75.70.Cn

# I. INTRODUCTION

Triggered by the success of graphene, the field of two-dimensional (2D) semiconductor (SC) crystals based on transition metal dichalcogenide (TMDC) monolayers encounters a spectacular development [1]. This new class of exciting materials presents several original characteristics. Their direct band gap [2] allows investigations by optical techniques, and the strong spin-orbit coupling combined with the lack of inversion symmetry result in non-equivalent valleys in the reciprocal space, that can be selectively addressed by circularly polarized light due to optical selection rules [3]. In a sense, the valley index could constitute a novel degree of freedom to carry and process information ("valleytronics" [4]). The exploration of the spin and valley degrees of freedom by all-optical experiments have been carried in the past years [5, 6, 7]. However, electrical spin injection yet remains elusive in these systems [8]. The ability to inject spin polarized currents would pave the way for new spintronic and spin-optronic devices [9,10,11], where one can imagine to take benefit from optical and spin properties of these materials. It could also benefit to future valleytronic devices, where electrical generation and control of the valley index is required. Due to the unique correlation between the spin and valley indices of charge carriers, this point is directly linked to electrical injection of spin polarized (and energy selected) carriers in TMDC. The problem of electrical injection into 2D semiconductors like $MoS_2$, $MoSe_2$, $WSe_2$, or $WS_2$ is thus a new challenge in this field, after the recent exploration of electrical spin injection into tridimensional (3D) semiconductors such as GaAs [9,10,11], Si [12], Ge [13], and in 2D Graphene [14]. The first experimental proof of electrical spin injection in 2D semiconductors has just been reported very recently: Ye *et al.* have shown that spin polarized holes can be injected in $WSe_2$ from the ferromagnetic semiconductor GaMnAs that acts as a spin aligner [15]. Unfortunately, this ferromagnetic material has a Curie temperature near 200K, far below room temperature. An alternative way consists in using ferromagnetic metals (FM) like Cobalt, Iron or Nickel, with Curie temperatures well above room temperature. In the context of the electrical spin injection into 3D semiconductors in the diffusive regime, it is well established that one has to overcome the impedance mismatch [16] between the FM and the SC layers: this is usually performed by inserting an oxide layer or a tunnel barrier [17]. The electrical spin injection from a ferromagnetic electrode into a single TMDC layer should, however, be totally different from the case of spin injection in a 3D semiconductor and the problem has to be reconsidered. In particular, the semiconductor nature of $MoS_2$ just below the metallic contact is questionable because covalent bonds could be formed between the metal and the TMDC monolayer, as discussed by Allain *et al.* [8]. Modification of the magnetic properties of $MoS_2$ below the contact, in particular a spin-polarization of the electron states near the Fermi level and induced magnetic moments, could also occur due to the direct contact between MoS$_2$ and the magnetic layer. First-principles calculations of the electronic structure based on the density functional theory (DFT) constitute a tool of choice to understand the bonding mechanisms between the FM and the TMDC layers and to give a clear description of the spin-polarized electron states at their interface.

Most of the first-principles studies on interfaces between metals and a single TMDC layer have focused on non-magnetic metals like Ir, Pd, Ru, In, Ti, Au, Mo or W [18,19]. Only few studies have concerned interfaces between Fe, Co or Ni and a TMDC monolayer: Dolui *et al.* have studied the Giant Magnetoresistance of Fe/MoS$_2$/Fe magnetic tunnel junctions and demonstrated that MoS$_2$ becomes conductive when the MoS$_2$ spacer only contains one or two layers, due to the strong interaction between Fe and S atoms; this behavior could, however, be due to the fact that Fe is present on both sides of the MoS$_2$ layer, and the case of a single Fe/MoS$_2$ interface has not been investigated there [20]. Considering an interface between a single Co atomic layer and a MoS$_2$ sheet, Chen *et al.* showed that the electronic structure of these two monolayers is drastically modified by a strong interface binding [21]; their results are however strongly influenced by the extreme thinness of the Co layer (which is so thin that it even becomes half-metallic), while Co electrodes in real Co/MoS$_2$-based devices would certainly be thicker and contain several Co atomic layers. Leong *et al.* have studied the Ni(111)/MoS$_2$ interface, but they mostly focus their study on the consequences of the insertion of a graphene layer between Ni and MoS$_2$, than on a detailed description of the Ni(111)/MoS$_2$ interface [22]. Finally, Yin *et al.* have recently calculated the electronic structure of the Fe/MoS$_2$ and Co/MoS$_2$ interfaces [23]. Their study of the Fe(111)/MoS$_2$ interface is based on a supercell in which a 3x3 slab of Fe(111) and a 4x4 cell of MoS$_2$ are stacked together. For Co/MoS$_2$, they used a supercell in which a MoS$_2$ monolayer is sandwiched between face-centered cubic (*fcc*) Co layers and an atomic structure which should not correspond to that of a real Co/MoS$_2$ interface, not only because Co actually crystallizes in the hexagonal compact (*hcp*) structure, but also because their multilayer is based on 4x4 Co(111) atomic layers superimposed on a 3x3 MoS$_2$ single layer: considering the experimental values of the lattice parameters (0.2507 nm for *hcp* Co and 0.312 nm for MoS$_2$), this interface corresponds to a relatively high lattice mismatch of 6.9%. Moreover, the MoS$_2$ layer would only be bound to Co on one of its two sides, in a realistic MoS$_2$/Co contact. For all these reasons, the genuine atomic structure of the Co/MoS$_2$ interface may probably be different from that used by these authors.

In the present paper, we considered supercells built from 5x5 Co(0001) atomic layers with the *hcp* stacking and a 4x4 MoS$_2$ single layer. These stacking would correspond to the very small lattice mismatch of 0.4% and would be more realistic for calculating the physical properties of the Co/MoS$_2$ interface between a Co magnetic electrode and a single MoS$_2$ layer.

After a brief description of the first-principles methods that we have used, we will first describe the atomic structure of the Co(0001)/MoS$_2$ interface, before giving details on the electronic states, magnetic moments, and charge transfers at the interface. We finally discuss the physical properties of the Co(0001)/MoS$_2$ interface in the perspective of spin injection in the 2D semiconductor MoS$_2$, and give insights on the possible utilization of the low strained Co/MoS$_2$ contact for in plane spin transport in lateral channels [17].

**II. FIRST-PRINCIPLES METHODS**

The ground state energy, the charge and the spin densities of all the supercells have been calculated self-consistently using the full-potential augmented plane waves + local orbitals (APW+lo) method implemented in the code WIEN2k [24]. The Kohn-Sham equation has been solved in the framework of the density functional theory (DFT), using the parametrization proposed by Perdew, Burke and Ernzerhof for the exchange and correlation potential that was treated within the generalized gradient approximation (GGA) [25]. In all our supercells, we used atomic sphere radii of 1.8, 1.8, and 2.0 atomic units (a. u.), respectively for Co, S and Mo atoms. The largest wave vector $K_{max}$ used for expanding the Kohn-Sham wave functions in the interstitial area between atomic spheres is given by the dimensionless parameter $R_{min}K_{max}$=6.0, where $R_{min}$ is the smallest atomic sphere radius of the supercell; this corresponds to an energy cut-off of 151 eV. The irreducible wedge of the Co/MoS$_2$ supercell two-dimensional Brillouin zone was sampled with a k-mesh of typically 24 different *k*-vectors, generated with a special *k*-grid used to perform Brillouin zone integrations with the modified tetrahedron integration method.

To model the Co/MoS$_2$ interface, we used rather big symmetric supercells consisting in a Co slab with *hcp* stacking and a thickness of five 5x5 (0001) monolayers (MLs), covered on each of its two sides by a 4x4 MoS$_2$ single layer, followed by vacuum. The thickness of the Co layer is sufficient to recover most of the electronic structure of bulk *hcp* Co at the center of the slab (shape of the density of states curves, value of the spin magnetic moment) and the thickness of the vacuum separation between periodically adjacent MoS$_2$ layers is above 1 nm, large enough to avoid interaction effects between them. We have chosen to use a periodic slab with a single MoS$_2$ layer on both sides of Co to get also identical surfaces on both sides of the Co slab and vacuum separation, which avoids artefacts such as charge transfer between surfaces across the Co layer. This kind of huge supercells contain 125 Co, 64 S and 32 Mo atoms and up the 36 non-equivalent atoms, depending on the relative positions of S and Co atoms at the interface. This high number of atoms is a necessary requirement to treat the problem of the realistic low-strained single MoS$_2$ contact that can be found, for example, at the source or the drain of a spin-field effect transistor (FET) based on a MoS$_2$ channel. The lattice parameter that we chose for the Co(5x5)/MoS$_2$(4x4) unit cell corresponds to four times the lattice parameter calculated for MoS$_2$ (0.319 nm).

The atomic structure of all the different MoS$_2$/Co(5MLs)/MoS$_2$ supercells that we have considered has been obtained by minimizing the forces acting on the different atoms. The minimum energy is achieved when all the atoms have reached their equilibrium position in the supercell. London dispersion forces are not included in the calculation.

Once the atomic structure has been calculated, we can choose to include spin-orbit coupling effects at each self-consistent loop, to check whether these effects modify the electronic structure of the Co(0001)/MoS$_2$ interface. This is performed within the second-order perturbation theory, for which we chose to include unoccupied electron states up to the maximum energy of 48 eV above the Fermi level, the magnetization being oriented along the Co(0001) axis.

### III. ATOMIC STRUCTURE OF THE Co(0001)/MoS$_2$ INTERFACE

We considered three different supercells, which correspond to three different manners of superimposing the 4x4 unit cell of 1H-MoS$_2$ (shown on Figure 1a) on the 5x5 unit cell of Co(0001) (Fig.1b). The first supercell (further labelled Supercell1) corresponds to the case where one of the interface S atoms of the 4x4 MoS$_2$ cell (for instance, the S atom at the corners of the 4x4 MoS$_2$ cell shown in Fig.1a) is located just above one of the *fcc* hollow atomic sites of the Co(0001) surface. The second supercell (Supercell2) corresponds to the case where one of the interface S atoms is on a top atomic site (*i. e.* just above an atom of the surface Co ML). We finally considered the last supercell (Supercell3), for which one of the interface S atoms is above one of the *hcp* hollow sites of the Co(0001) surface (*i. e.* just above an atom of the subsurface Co ML). These three supercells are represented in Figure 2. The atomic structure of these three supercells has been calculated self-consistently without including spin-orbit coupling. After all the atoms have reached their equilibrium position, we observed that the ground state energy is the lowest for Supercell1, and respectively 0.0827 eV and 0.813 eV higher for Supercell2 and Supercell3. These energy differences correspond to the whole supercells, which all contain 2 MoS$_2$/Co interfaces with 16 MoS$_2$ formula units in the 4x4 MoS$_2$ cell. Consequently, the difference per MoS$_2$ formula unit, between the ground state energies of the different supercells is only of 2.6 meV between Supercell2 and Supercell1, and of 25.4 meV between Supercell3 and Supercell1. From now, we will only consider the physical properties of Supercell1, which corresponds to the lowest energy interface structure.

Each of the two Co/MoS$_2$ interfaces in the supercell involves 25 Co and 16 S atoms. One of these 16 interface S atoms (further labelled S3) is covalently bound to 3 different but equivalent interface Co atoms (labelled Co3); 3 of the 16 interface S atoms (labelled S2) are each bound to 2 different but equivalent Co atoms (Co2), and all the 12 remaining interface S atoms are bound to a single Co atom. The position of all these atoms of the Co/$MoS_2$ interface is shown in Figure 3. The distance between interface S and Co atoms is of 0.236 nm between S3 and Co3, 0.234 nm between S2 and Co2, and of 0.221 or 0.222 nm between the 12 remaining S and Co interface atoms. These interatomic distances are very close to those which have been measured for bulk cobalt sulfides (0.232 nm for $CoS_2$ with the pyrite structure [26]). This clearly shows that bonding between the $MoS_2$ single layer and the Co(0001) surface has a covalent nature and is not due to the Van der Walls interaction. Each of the interface Co atoms are bound to a single S atom, except the four ones marked by a star in Fig.3.

The atomic layers present a small warping near the Co/$MoS_2$ interface, due to the different numbers of chemical bounds formed by the non-equivalent interface S atoms. Based on the averaged values of the z-coordinates (the z-axis being perpendicular to the interface) calculated for the atoms in the different atomic layers, we can estimate the average distance between successive atomic layers: we obtain an average distance of 0.205 nm between the interface Co and interface S atomic layers, and average distances of 0.160 and 0.154 nm between the Mo layer and the interface S and external S layers, respectively. These later distances can be compared to the distance of 0.157 nm calculated between the S and Mo atomic layers of an isolated $MoS_2$ sheet. The undulation of the successive atomic layers (difference between the highest and lowest z-coordinates) is respectively of 0.011 nm, 0.029 nm, 0.026 nm and 0.023 nm for the interface Co, interface S, Mo and external S atomic layers.

## IV. ELECTRONIC STRUCTURE OF THE Co(0001)/$MoS_2$ INTERFACE

The majority and minority spin band structures of the Co/$MoS_2$ slab are shown in Figures 4b and 4c. They contain a huge number of bands, some of them corresponding to cobalt bands and others to $MoS_2$ bands, folded on themselves, with important band gap opening at the center and the edges of the two-dimensional Brillouin zone. The shaded areas which appear in these band structures correspond to the Co *d*-band continuum (below -0.5 eV for majority spin and across the Fermi level for minority spin electrons). Additional bands, which do not correspond to the folded bands of Co or $MoS_2$ also appear in this figure; this is for instance the case between -0.5 eV and 0.5 eV for majority spin electrons. These new bands correspond to interface Bloch states involving covalently bound Co and S atoms. They strongly modify the physical properties of $MoS_2$, giving a metallic behavior to this layer, induced by interface covalent bonding. Some of these interface bands have a non-negligible dispersion near the Fermi level (electrons in the metallic phase of $MoS_2$ at the Co/$MoS_2$ interface will not have a high effective mass).

Each energy band in Figs.4b and 4c is drawn with small circles, the radius of which is proportional to the contribution of the MoS$_2$ layer to the corresponding Bloch states. The band of the Co/MoS$_2$ slab corresponding to the bottom of the conduction band of the MoS$_2$ single layer appears near 0.32 eV above the Fermi level (it is indicated with a blue arrow on Fiq.4b). This band is more clearly visible for minority spin (Fig.4b) than for majority spin electrons for which it is located in the continuum of Co *d*-bands (Fig.4c). The difference between the energy of this band and the Fermi level corresponds to the height of the Schottky barrier [19], the value of which can be estimated at $\phi_B = 0.32$ eV for the Co/MoS$_2$ interface. Our calculation of the Schottky barrier height is performed at the DFT level. As discussed by Zhong *et al.* [27], this seems to give results in a better agreement with experimental ones than those computed with the GW method, probably due to the fact that many-electron effects are greatly depressed by the charge transfer at the MoS$_2$/metal interface, which significantly screens electron-electron interaction [27].

Figure 5 shows the density of states (DOS) of the MoS$_2$ layer at the Co(0001)/MoS$_2$ interface. This DOS curve is continuous between -8 eV and energies well above the Fermi level, confirming the metallic character of the MoS$_2$ layer when it is covalently bound to the cobalt surface. This holds for majority as well as for minority spin electrons. Despite the strong modification of the electronic structure of MoS$_2$, important DOS peaks can be identified in Fig.5 as belonging to the valence and conduction bands of the isolated MoS$_2$ layer: The DOS curves of the isolated MoS$_2$ sheet has also been represented in Fig.5, where it has been shifted to coincide to the main DOS peaks of MoS$_2$ at the Co/MoS$_2$ interface. The bottom of the conduction band in the shifted DOS curve of the isolated MoS$_2$ layer is near 0.32 eV above the Fermi level, which confirms the value of the Schottky barrier height estimated from the band structure of the supercell. This estimation of the Schottky barrier height is larger than the values measured experimentally (between 60 meV in [28] and 121 meV [29]). These measurements are however performed on exfoliated MoS$_2$ flakes. In these cases, the flakes are exposed to air before Co is deposited. It means that impureties and molecules can be trapped at the interface, resulting in localized states within the gap that can modify the pinning of the fermi level, and thus the Schottky barrier height. It would be interesting to compare our result with a full epitaxial Co/MoS$_2$ hybrid system elaborated in ultra-high vacuum, when it will be available. Note that our estimation of the Schottky barrier height of the Co/MoS$_2$ interface is comparable to values found in previous studies for similar interfaces involving another metal, like the Ti/MoS$_2$ system [19].

The majority and minority spin densities of states give access to the spin polarization $P_S$ near the Fermi level in the MoS$_2$ layer. This important quantity, defined as the difference between the majority and the minority spin densities of states divided by their sum, clearly indicates if a spin-polarized current can be injected through the Co/MoS$_2$ interface to a MoS$_2$ channel. DOS curves shown in Fig.5 have been used to calculate the spin-polarization represented in Figure 6: It takes a negative value $P_S = -31\%$ at the Fermi level. This value decreases however to $P_S = -16\%$ when spin-orbit coupling is taken into account. Even if the spin-polarization at the Fermi level is lower in the MoS$_2$ layer bound to the Co surface than in pure cobalt, it should be high enough to perform spin-injection in MoS$_2$ from a Co/MoS$_2$ contact. The estimation of the spin polarization at the Fermi level for Co/MoS$_2$ interface is of importance for spin injection. As pointed out by Mazin [30], an accurate determination of the electric current spin polarization would however require complementary DFT-based calculations (including the transmittance of the whole complex structure and the effects of the bias voltage).

**V. SPIN MAGNETIC MOMENTS AT THE Co(0001)/MoS$_2$ INTERFACE**

The spin magnetic moment of Co atoms is on average 8% lower at the Co(0001)/MoS$_2$ interface than in bulk *hcp* Co (1.69 μ$_B$), with values that depend on the kind of interface S atom to which they are covalently bound: The interface Co atoms that show the highest spin magnetic moment (1.66 μ$_B$ and 1.63 μ$_B$) are the 4 atoms which are not bound to S atoms, followed by Co$_3$ (1.62 μ$_B$) and Co$_2$ atoms (1.57 μ$_B$). All the other interface Co atoms have a spin-magnetic moment between 1.48 and 1.50 μ$_B$: the lowering of the interface Co atom magnetic moment is more important when Co atoms are more strongly bound to S atoms, with a shorter Co-S bond length. All the interface S atoms have a small spin magnetic moment with the same sign as the Co atom magnetic moments, with values between 0.012 and 0.016 μ$_B$. The spin magnetic moment of S atoms in the external S layer is even smaller (0.003 to 0.004 μ$_B$). The spin magnetic moment of Mo atoms (between -0.029 and -0.024 μ$_B$) is antiferromagnetically coupled to the Co and S magnetic moments, except when these Mo atoms are bound to one or to two S$_2$ atoms (in this case, Mo spin magnetic moments respectively take the positive values 0.008 and 0.050 μ$_B$).

**VI. CHARGE TRANSFER AT THE Co(0001)/MoS$_2$ INTERFACE**

To calculate the charge transfer between atoms induced by covalent bonding at the Co(0001)/MoS$_2$ interface, we proceeded as follows: first, we calculated the majority spin electron density $n_\uparrow(\mathbf{r})$. In a second step, we calculated the majority spin density $n_{\uparrow,Co}(\mathbf{r})$ when all the Co atoms keep exactly the same positions as in Supercell1, while Mo and S atoms have been removed. In a third step, we obtained the majority spin density $n_{\uparrow,MoS_2}(\mathbf{r})$, calculated when Mo and S atoms keep the same positions as in Supercell1 and Co atoms have all been removed. The majority spin space-dependent charge transfer can finally be obtained from $\Delta n_\uparrow(\mathbf{r}) = n_\uparrow(\mathbf{r}) - \{n_{\uparrow,Co}(\mathbf{r}) + n_{\uparrow,MoS_2}(\mathbf{r})\}$. The minority spin charge transfer is similarly given by $\Delta n_\downarrow(\mathbf{r}) = n_\downarrow(\mathbf{r}) - \{n_{\downarrow,Co}(\mathbf{r}) + n_{\downarrow,MoS_2}(\mathbf{r})\}$. Figure 7 shows a top and a side view (the direction of observation corresponds to the blue arrow in Fig. 3) of the calculated three-dimensional majority and minority spin charge transfers $\Delta n_\uparrow(\mathbf{r})$ and $\Delta n_\downarrow(\mathbf{r})$. Red and green areas respectively correspond to a local excess and a local lack of electrons. As expected, we see on this figure that charge transfers occur between interface Co and S atoms along the Co-S covalent bonds. This figure shows that the reduction of the Co magnetic moment at the interface is due to an excess of minority spin electrons, concomitant with a lack of majority spin electrons on interface Co atoms. Similarly, the magnetic moment of Mo atoms is due to an excess of majority spin and a lack of minority spin electrons for the Mo atoms which have a positive spin magnetic moment (right hand side of Figs.7b and 7d), and mostly to an excess of minority spin electrons for Mo atoms which have a negative magnetic moment (left hand side of Fig.7d).

## VII. PERSPECTIVES: TOWARDS ELECTRICAL SPIN INJECTION IN MoS$_2$

In the context of spin injection into bulk semiconductor materials, it has been established that the so-called conductivity mismatch [16] between the ferromagnetic metal injector and the semiconductor constitutes a fundamental obstacle for efficient spin injection at the ferromagnetic metal/semiconductor interface in the diffusive regime. To circumvent this problem, it has been shown that a thin tunnel barrier must be introduced between FM and SC, inducing an effective spin dependent resistance [17]. A possible tunnel barrier that can be exploited is the natural Schottky barrier that occurs between the two materials. As spin injection in tunnel regime is desirable, a careful engineering of the doping in the semiconductor close to the interface is required in order to make the Schottky barrier thin enough to behave as a tunnel barrier. It has been realized successfully in Fe/GaAs system [10] where GaAs was gradually n-doped up to $10^{19}$ cm$^{-3}$ at the interface, resulting in a very efficient electrical spin injection from Fe into GaAs in tunnel regime.

In the context of TMDCs, the general problem of electrical contacts on TMDCs for transport in a two-dimensionnal channel is a challenging task (tackled for example by phase engineering techniques [31,32], as well as using contacts based on heterostructures involving graphene [22]. Concerning spin injection with Co/MoS$_2$ (or other FM/TMDC [33]) interfaces, one has to consider the Schottky barrier between the metallic phase of MoS$_2$ just below the contact, labeled hereafter (MoS$_2$)*, and the semiconductor MoS$_2$ channel out of the contact. As in GaAs, one could imagine to strongly increase the doping in MoS$_2$ close to the (MoS$_2$)*/MoS$_2$ one-dimensional border, see Figure 8. Even if, up to now, spatially controlled chemical doping of a single layer TMDC is still a challenge, such an in-plane localized doping has been already obtained with the help of additional gate electrodes developed successfully for in-plane P-I-N junctions in TMDC Light Emitting Diodes [34]. Considering the electrical spin injection tunnel process at this (MoS$_2$)*/MoS$_2$ (n-doped) one-dimensional interface, one has to compare, for majority and minority spin, the compatibility of the symmetries of the electron wave functions in (MoS$_2$)* at the Fermi level, with the ones in the conduction band of the MoS$_2$ single layer semiconductor channel, in order to estimate the efficiency of the tunneling process. The conduction electron wavefunctions in the isolated MoS$_2$ channel exhibit a strong Mo $d_z^2$ character (as well as a smaller Mo $s$ character), and a S ($p_x+p_y$) character. The DOS at the Fermi level calculated for (MoS$_2$)* is qualitatively different from the DOS at the bottom of the conduction band of MoS$_2$. Due to the strong hybridization between Co and S atomic orbitals at the Co/MoS$_2$ interface, Bloch electron states at the Fermi level result from a non-trivial combination of atomic orbitals that involve Mo-$d_z^2$, S-($p_x+p_y$) and also other orbitals, see Table 1. However, the contribution of Mo-$d_z^2$ and S-($p_x+p_y$) atomic orbitals is not negligible and still represents 33% and 37% of the total DOS of (MoS$_2$)* at the Fermi level, respectively for majority and minority spin electrons. The symmetry of the majority and minority spin electron states at the Fermi level in (MoS$_2$)* is thus partly compatible with the symmetry of Bloch states in the conduction band in the MoS$_2$ channel.

The efficiency of electrical injection in the MoS$_2$ channel in the tunnel regime should also depend on the direction of the one dimensional border between (MoS$_2$)* and MoS$_2$. We know that the Bloch vector of electrons in the conduction band of the MoS$_2$ channel, after tunneling through the Schottky barrier, corresponds to one of the $K$-valleys. We also know, using the simple model of a free electron with mass $m$ and energy $E$ propagating in a plane towards a straight one-dimensional potential step where the potential jumps from 0 to $U > E$, that the penetration depth that characterizes the exponential decay of the wave function after the step is given by $\left\{\dfrac{2m}{\hbar^2}\left(U - E + \dfrac{\hbar^2 k_{//}^2}{2m}\right)\right\}^{-\frac{1}{2}}$; $k_{//}$ is the component of the electron two-dimensional Bloch vector parallel to the straight border, and $\hbar = h/2\pi$ is Planck's constant. The penetration depth is thus maximum for $k_{//} = 0$. It follows that the straight (MoS$_2$)*/MoS$_2$ border should be perpendicular to the ($\Gamma K$) direction of the two-dimensional Brillouin zone, in order to get a spin-polarized current with maximum intensity reaching a $K$-valley in the MoS$_2$ channel. In other words, the one dimensional border between (MoS$_2$)* and MoS$_2$ should correspond to an armchair-like border, as shown in Fig.8a.

All the points considered in this section should have an impact on the intensity and on the spin polarization of the electron current injected into MoS$_2$, but a full calculation of the spin transport process by DFT devoted tools, in particular of the transmittance of a realistic Co/(MoS$_2$)*/MoS$_2$/(MoS$_2$)*/Co device, is a non-equilibrium problem clearly beyond the scope of this paper.

## VIII. CONCLUSIONS

In this paper we have investigated the electronic structure of a single low-strained *hcp* Co(0001)/MoS$_2$ interface, using first-principles calculations based on the functional density theory, in order to estimate the potentiality of a cobalt injector for electrical spin injection into a MoS$_2$ monolayer, in view of spintronic devices as spin-FET. We first described the lowest energy atomic structure and show that interface S atoms are covalently bound to one, two or three interface Co atoms. A lower spin magnetic moment is observed for Co atoms at the interface, together with a small magnetization of S atoms. Mo atoms also hold small magnetic moments which can takes positive as well as negative values. The induced magnetic moments have been interpreted in terms of majority and minority spin charge transfers at the interface. Band structures and density of states curves have been calculated for minority and majority spin electrons in the hybrid system, taking into account spin-orbit coupling. We demonstrate that MoS$_2$ just below the cobalt contact becomes metallic due to hybridization with Co *d* orbitals. For this metallic phase of MoS$_2$, labelled (MoS$_2$)*, a spin polarization of -16 % is calculated at the Fermi level and the Schottky barrier height of the (MoS$_2$)*/MoS$_2$ contact has been estimated around 0.32 eV. Finally, the symmetry of the electron wave functions at the Fermi level in the metallic phase of MoS$_2$ just below the Co contact and the shape (armchair-like) of the one-dimensional border between (MoS$_2$)* and MoS$_2$ have been considered and their impact on spin injection into the MoS$_2$ channel have been discussed. Further first-principles calculations, including the effect of the applied bias voltage and the transmittance of the channel and of the Co/(MoS$_2$)*/MoS$_2$ borders, should be led in order to calculate the non-equilibrium spin-dependent transport phenomena of MoS$_2$-based devices with Co electrodes.

## ACKNOWLEDGEMENTS

The authors acknowledge the Université Fédérale de Toulouse-Midi-Pyrénées and Région Midi-Pyrénées for the PhD grant SEISMES as well as the grant NEXT n° ANR-10-LABX-0037 in the framework of the « Programme des Investissements d'Avenir". This work

**TABLE CAPTIONS:**

**Table 1:** Total and partial (for the most important atomic orbitals) Mo and S majority and minority spin density of states at the Fermi level for the Co/MoS$_2$ interface. All the results are given in the same arbitrary unit.

**FIGURE CAPTIONS**

**Figure 1:** Atomic structure (top views) of (a): the 4x4 unit cell of $MoS_2$, and (b): the 5x5 unit cell of a 5MLs Co(0001) slab.

**Figure 2:** Side views of the atomic structure of the $Co/MoS_2$ supercells corresponding to (a): Supercell1 (b): Supercell2, and (c): Supercell3.

**Figure 3:** Top view of the atomic structure of the $Co(0001)/MoS_2$ interface for Supercell1. The atoms Co2 (dark blue spheres), Co3 (medium blue), S2 (red) and S3 (orange) discussed in section III are indicated. The Co atoms marked by a star are those which are not directly bound to a S atom.

**Figure 4:** Band structure of (a): the 4x4 $MoS_2$ single layer supercell, (b): the $Co/MoS_2$ slab for majority spin, (c): the $Co/MoS_2$ slab for minority spin, (d): the 5x5 Co(0001) slab for majority spin and (e): the 5x5 Co(0001) slab for minority spin. The radius of the red circles in panels (b) and (c) is proportional to the $MoS_2$ contribution to the electron states. The blue arrow in panel (a) indicates the minimum of the conduction band in $MoS_2$; its corresponding position in the band structure of the $Co/MoS_2$ slab is also shown with a blue arrow in panel (b), where it corresponds to the height of the Schottky barrier.

**Figure 5:** Contribution of a $MoS_2$ monolayer to the density of states of Supercell1 (dark curves). The density of states of an isolated $MoS_2$ layer is also represented after an energy shift (red curves). The upper and lower parts of the figure respectively correspond to majority and minority spin electrons.

**Figure 6:** Spin polarization in the $MoS_2$ layer, calculated without (red curve) and with (dark curve) spin-orbit coupling.

**Figure 7:** (a): Top view and (b): side view of the majority spin charge transfer $\Delta n_\uparrow(\mathbf{r})$ at the interfaces of the slab. (c): Top view and (d): side view of the minority spin charge transfer $\Delta n_\downarrow(\mathbf{r})$ at the interface. Positive and negative values are respectively shown with red and green colors. Yellow, light purple and blue spheres (when visible), display the position of S, Mo and Co atoms, respectively**.**

**Figure 8:** Sketch representing (a): the bottom view (below the Co contact) and (b): the side view of the physical area involving the Co(0001) contact, $(MoS_2)^*$, the Schottky barrier $(MoS_2)^*/MoS_2$ obtained by suitable doping and the $MoS_2$ channel. The one-dimensional border which should be designed to lower the exponential decay of wave functions in the Schottky barrier is indicated by red a dashed line in (a), together with the ($\Gamma K$) direction of $MoS_2$ (red arrow). The corresponding Schottky barrier profile is represented in panel (c).

# TABLES:

**TABLE 1:**

| Contribution | Majority spin | Minority spin |
|---|---|---|
| Mo: *total* | 0.304 | 0.542 |
| Mo: *s+$d_z^2$* | 0.104 | 0.182 |
| S: *total* | 0.127 | 0.228 |
| S: *$p_x+p_y$* | 0.037 | 0.106 |
| $MoS_2$: total | 0.431 | 0.770 |

**FIGURES:**

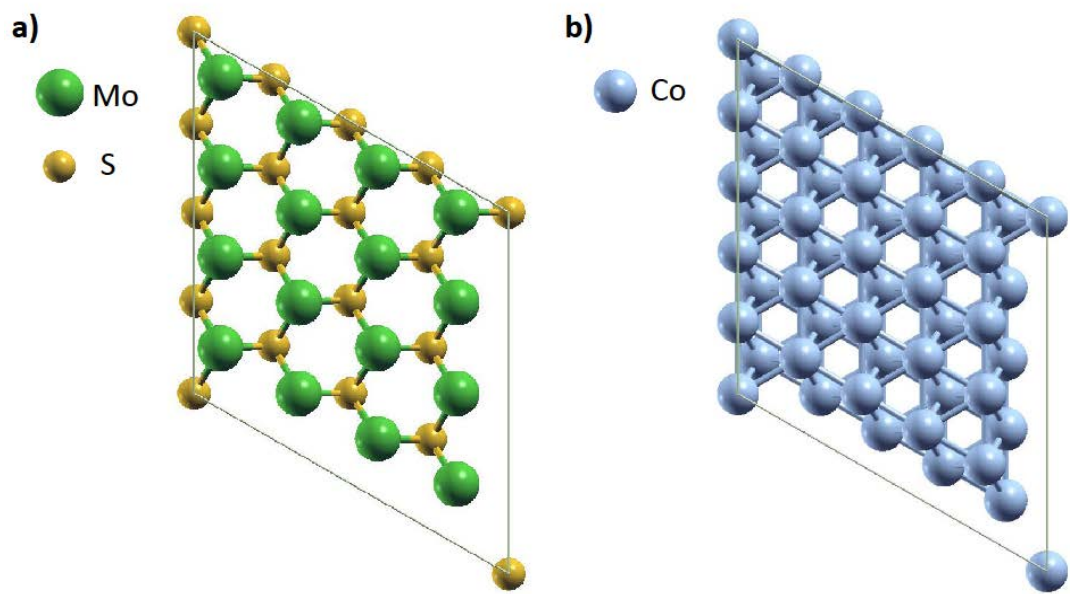

**Figure 1**

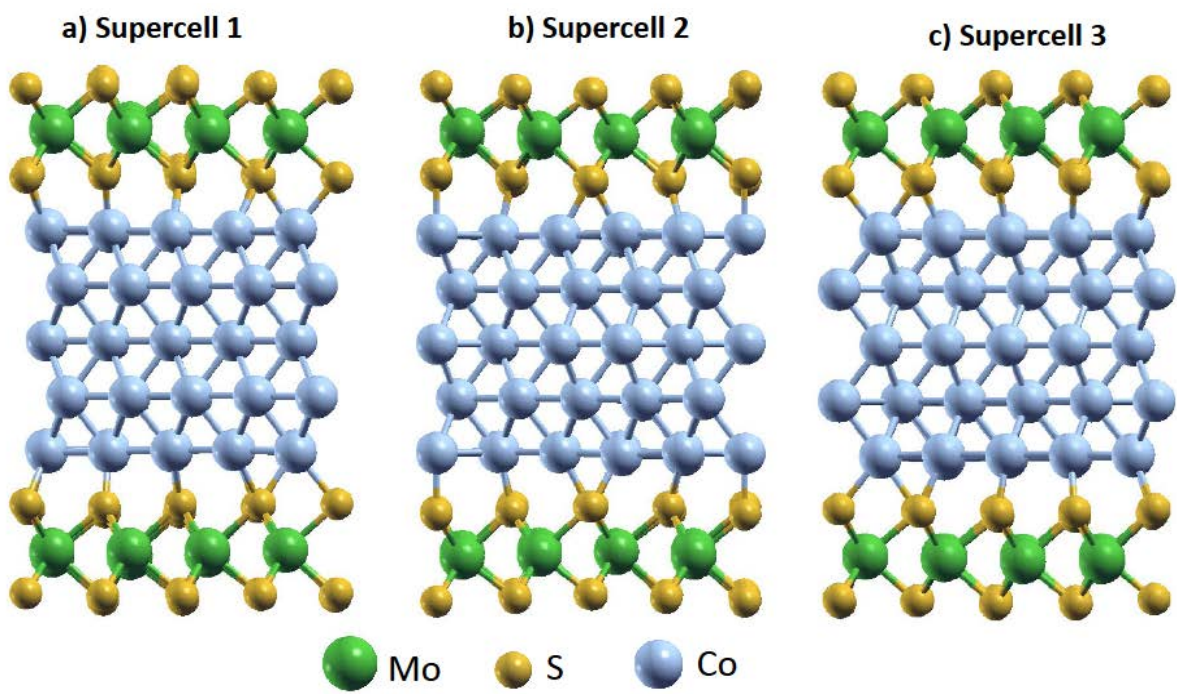

**Figure 2**

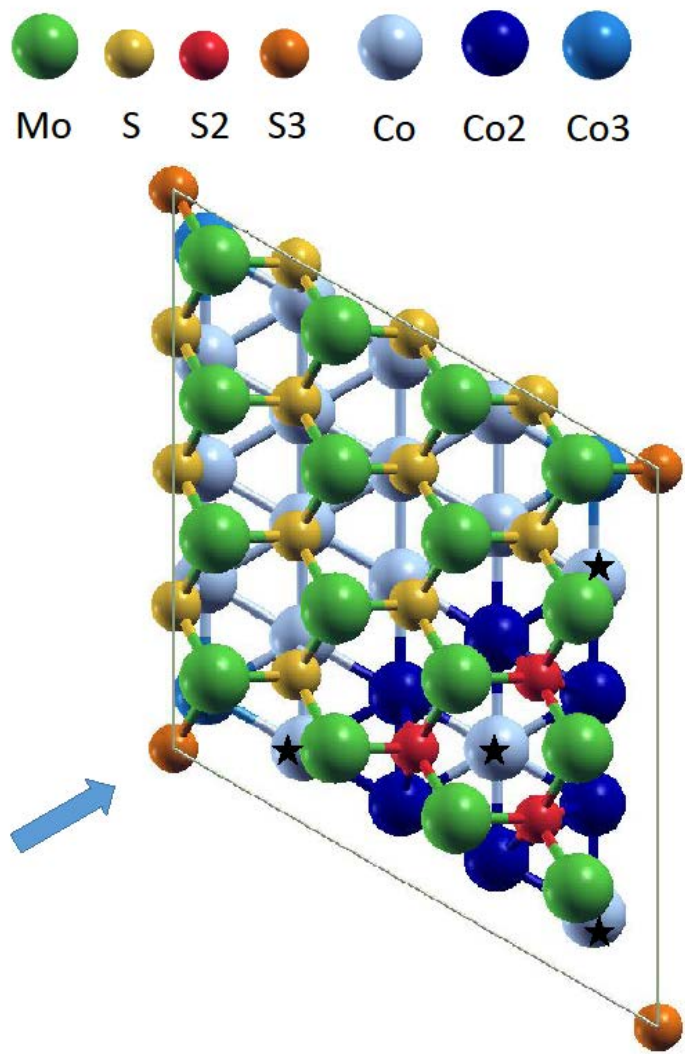

**Figure 3**

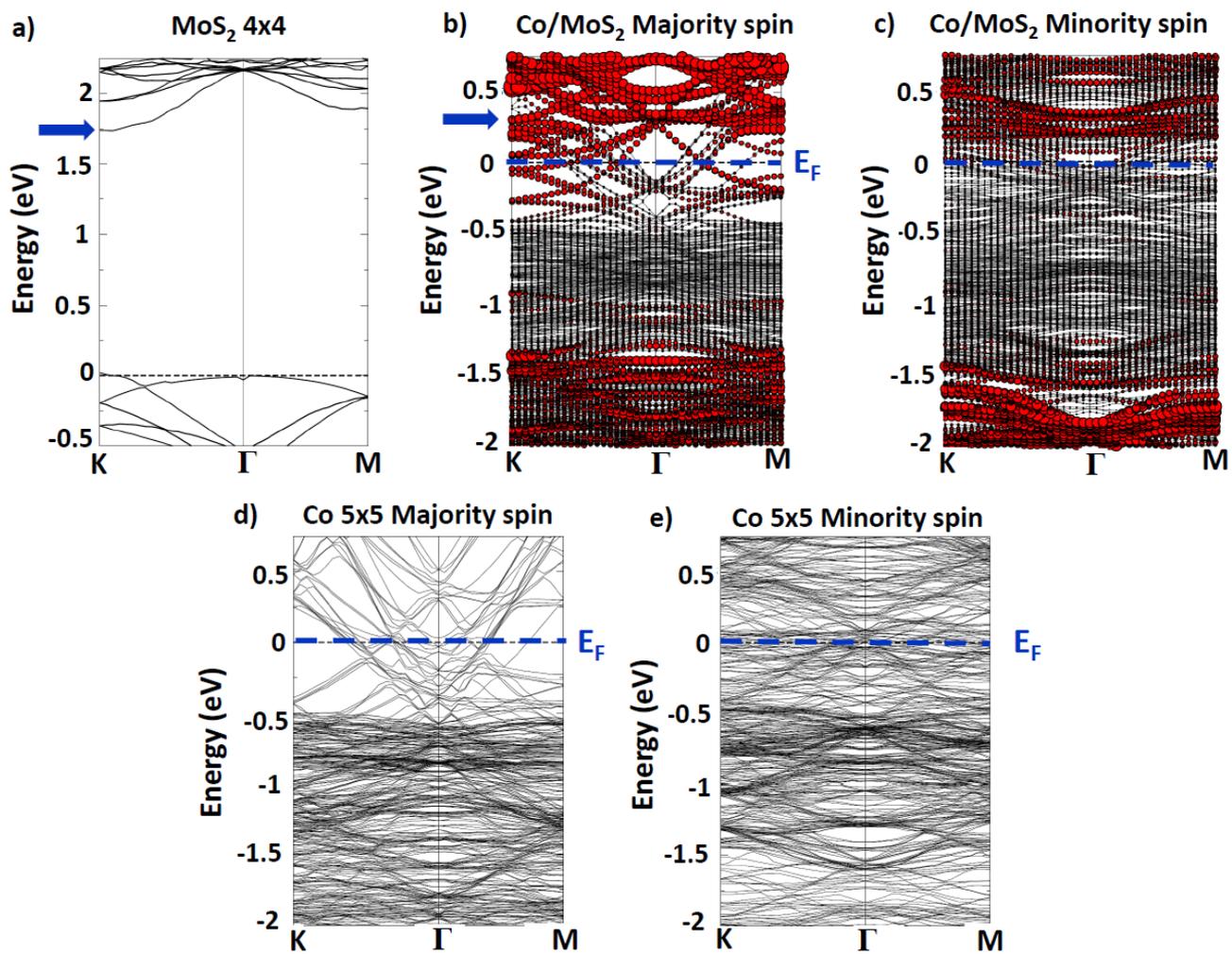

**Figure 4**

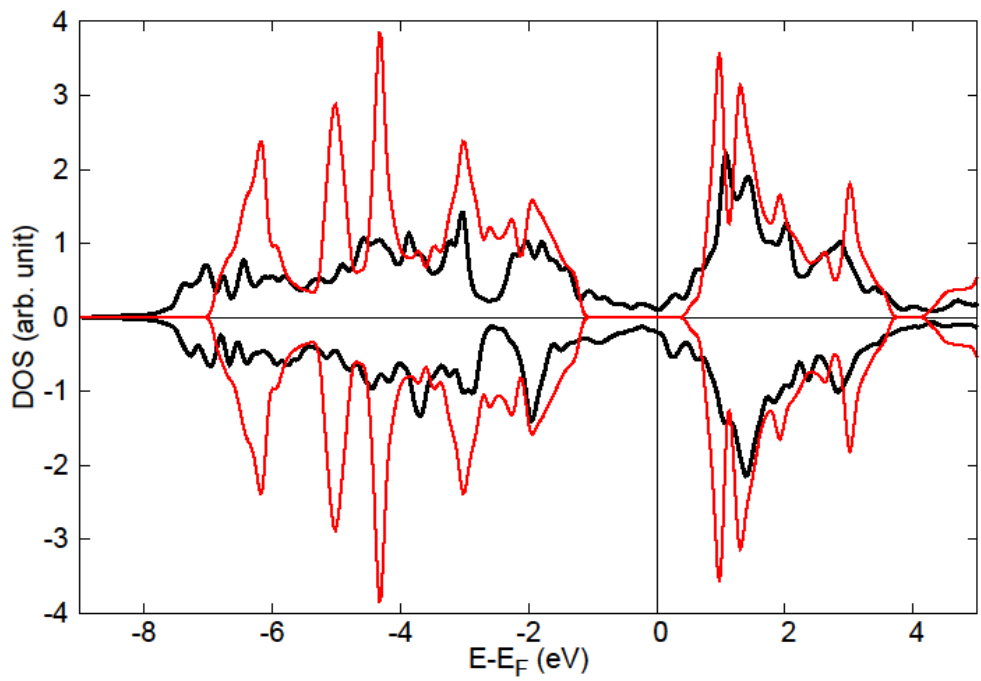

**Figure 5**

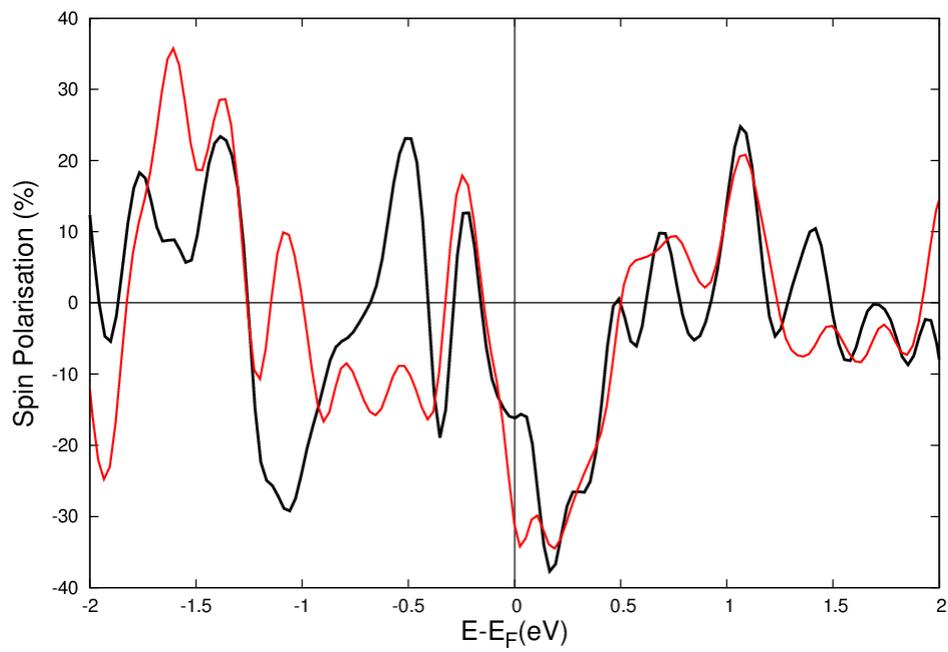

**Figure 6**

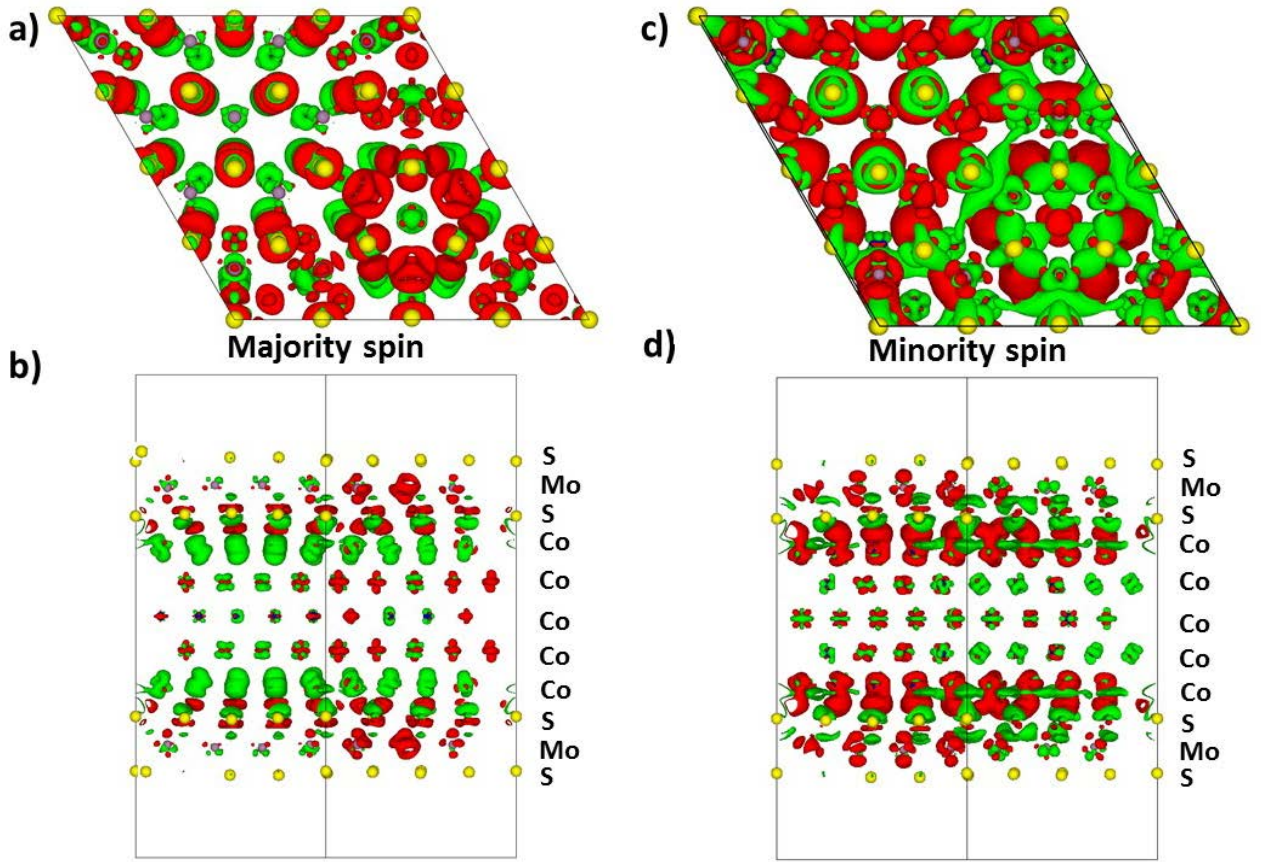

**Figure 7**

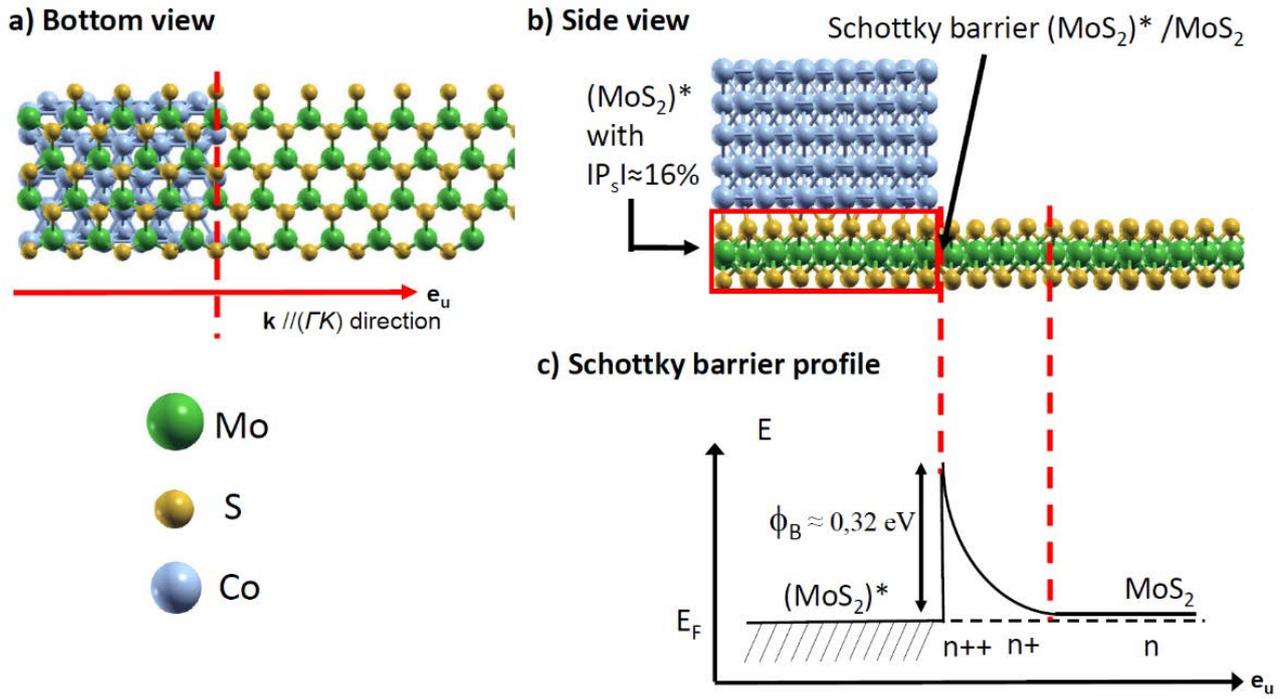

**Figure 8**